\documentclass[aps,prl,twocolumn,showpacs,amsfonts,superscriptaddress]{revtex4-1}

\usepackage{amsmath}
\usepackage{amssymb}
\usepackage{dsfont}
\usepackage{graphicx}
\usepackage{multimedia}
\usepackage{color}
\usepackage{xcolor}
\usepackage{ bbold }
\usepackage{bm}
\usepackage{bbm}

\begin{document}

\title{Correspondence between bulk equilibrium spin-currents and edge spin
accumulation in wires  with spin-orbit coupling}

\author{I. V. Tokatly} 
\email{ilya.tokatly@ehu.es}
\affiliation{Nano-Bio Spectroscopy Group, Departamento  F\'isica de Materiales, 
Universidad del Pa\'is Vasco, Av. Tolosa 72, E-20018 San Sebasti\'an, Spain} 
\affiliation{IKERBASQUE, Basque Foundation for Science, E-48011 Bilbao, Spain}
\affiliation{Donostia International Physics Center (DIPC) - Manuel de Lardizabal 5, E-20018 San Sebasti\'{a}n, Spain}

\author{B. Bujnowski}
\email{bogusz.bujnowski@gmail.com}
\affiliation{Donostia International Physics Center (DIPC) - Manuel de Lardizabal 5, E-20018 San Sebasti\'{a}n, Spain}

\author{F. S. Bergeret}
\email{fs.bergeret@csic.es}
\affiliation{Centro de F\'isica de Materiales (CFM-MPC) Centro Mixto CSIC-UPV/EHU - 20018 Donostia-San Sebastian, Basque Country, Spain}
\affiliation{Donostia International Physics Center (DIPC) - Manuel de Lardizabal 5, E-20018 San Sebasti\'{a}n, Spain}

\date{\today}

\begin{abstract}  
We demonstrate that the interplay of Zeeman and spin-orbit coupling fields in a 1D wire leads to an equilibrium spin current that manifests itself in a spin accumulation at the wire ends with a polarization perpendicular to both fields. This is a universal property that occurs in the normal and superconducting state independently of the degree of disorder. We find that the edge spin polarization transverse to the Zeeman field is strongly enhanced in the superconducting state when the Zeeman energy is of the order of the superconducting gap. {This 
 hitherto unknown  transverse magnetic susceptibility  can be much larger than the longitudinal  one, and  generalises the well established theory of the Knight-shift in superconductors to the case of finite systems. }
\end{abstract}


\maketitle

Spin currents have been a subject of intensive investigations in several branches of condensed matter physics \cite{Fabian2004,Rashba2007,Maekawa2012,Eschrig2011}.
Special  attention was brought to equilibrium spin currents (ESC) that may exist in materials with spin-orbit coupling (SOC). The interpretation of the ESC in such systems remained under debate for a  while  because in the presence of SOC the spin is not conserved in a customary sense \cite{Rashba2007,Shi2006,EngelRashbaHalperin2007,Rashba2003}. 
The theoretical controversy can be removed by treating the SOC as an external SU(2) gauge field \cite{Tokatly2008,Tokatly2010}, but from the experimental point of view ESC still remain elusive. There were several suggestions to detect ESC indirectly, for example in form of resulting mechanical and spin torques \cite{Sonin2007a,Sonin2007}. However,  to best of our knowledge such measurements have not been achieved and the ESC remains to be elucidated. 

In this letter, we demonstrate a correspondence between ESC in wires 
with SOC and a transverse spin accumulation at the edges of the corresponding finite system.
This correspondence is universal in the sense that it holds for any many-body one-dimensional system, provided the particle-particle interaction is spin-independent. ESC appear when a Zeeman field $h$ with a component perpendicular to the SOC is applied. The bulk ESC is always accompanied with the edge spin accumulation that is transverse to both the Zeeman field and SOC, and as a function of $h$  shows a cusp when  $h=\sqrt{\mu^2+\Delta^2}$, where $\mu$ is the chemical potential of the wire and $\Delta$ the induced superconducting gap.  
Interestingly, this cusp shows up in both  the normal ($\Delta=0$) \cite{Dolcini2018} and superconducting state. 
More striking, when $\Delta\ll \mu$,  the transverse spin accumulation shows a sharp maximum at $h\approx\Delta$ which can be much larger than the magnitude of the spin accumulation at the cusp.  We analyze in detail this maximum of the spin accumulation and show its robustness against disorder. Finally we present analytical results for the spatial distribution of the magnetic moment induced as a response to the Zeeman field.  We find that the transverse susceptibility close to the edge of the wire can be much larger than the longitudinal one, provided the SOC is small. This generalizes  the well known Knight-shift in bulk systems for the case of finite samples \cite{abrikosov1962spin,Abrikosov1988, Gorkov2001}.

In the context of Majorana fermions \cite{Sau2010,Oreg2010,Mourik2012,Das2012,Deng2012,Hansen2016,Sticlet2012a,Ojanen2012}
the presence of a transverse polarization has been identified as a signature of the  topological transition \cite{Sticlet2012a,Szumniak2017,Bjornson2015} when $h=\sqrt{\mu^2+\Delta^2}$.  However, as discussed in Ref.~\onlinecite{Dolcini2018} such transverse polarization also appears in the normal state
and therefore this identification has to be treated with care. Our work shows that the transverse spin polarization is a universal property of any one-dimensional system supporting ESC and it exists at all values of $h$, including those far below the topological transition.   
 


We consider a 1D system with a linear in momentum  SOC in the presence of a Zeeman or exchange field. As we will show, an equilibrium spin current, appears when the Zeeman field vector is perpendicular to the SOC field. Without loss of generality let the wire be oriented along the x-axis, Rashba SOC along the y axis and a Zeeman field along the z-axis. The many-body Hamiltonain in that case reads
\begin{flalign}
H=\sum_{i=1}^N\left[\frac{\hat{p}_{x_i}^{2}}{2m}+\alpha\hat{p}_{x_i}\sigma_i^y+h\sigma_{i}^z+V_{c}(x_i)\right] +
H_{int}\;,
\label{Eq01_Hamiltonian}
\end{flalign}
where the index $i$ labels the particles, $V_c(x)$ is a confinement potential, $\hat{p}_{x_i}=-i \partial_{x_i}$ is the momentum operator, $\alpha$ is the strength of the SOC, $h$ is the Zeeman field and ${H}_{int}$ describes a spin-independent interaction. 
This Hamiltonian is real and therefore its eigenfunctions can be chosen real. This implies that in equilibrium there is no component of the spin density parallel to the SOC, $s^y(x)=0$.  Similarly, by defining the spin current operator in the standard way \cite{Rashba2007,EngelRashbaHalperin2007,Sonin2007a,Tokatly2008},
\begin{equation}
 \hat{J}_x^a(x)=
\frac{1}{4}\sum_{i=1}^N\Big\{\Big\{\frac{\partial H}{\partial\hat{p}_{x_i}},\sigma_i^a\Big\},\delta(x-x_i)\Big\},
\end{equation}
we find that the system can only support an equilibrium spin current ${J}_x^y(x)$ polarized along the $y$-axis.


In equilibrium the equation of motion for the spin density takes the standard form,
\begin{flalign}
\partial_{x}{J}^{y}_x(x)+hs^{x}(x)=0\; ,\label{eq:continuity}
\end{flalign}
where the last term is the usual spin torque due to the Zeeman field.  This equation shows that ESC in the bulk of the system may exist only if the $x$-component of the spin is accumulated at the edges. In other words, Eq.~(\ref{eq:continuity}) establishes a correspondence between the bulk property of the system, the ESC, and the net spin accumulation transverse to the Zeeman field at the edge of a finite system. 
{An example of this correspondence has been  obtained   in Ref.\cite{Dolcini2018} for  a non-interacting finite 1D wire contacted to leads. }

{ In a  semi-infinite wire}, where all particles are confined in the right half-space, the spin accumulation at the left edge is obtained by integrating Eq.(\ref{eq:continuity})
\begin{equation}
S^{x}\equiv\int_{-\infty}^{+\infty}s^{x}(x)=-\frac{1}{h}{ J}_x^{y}(\infty)\;.\label{eq:S_accumulation}
\end{equation}
Thus the edge spin accumulation can be determined by calculating the ESC which is a bulk property independent of boundary conditions. For a finite wire, the transverse spin accumulation at opposite edges is equal in magnitude but oriented in the opposite direction, as shown schematically in Fig. \ref{Fig:scheme}a.  In the following we will use this relation to compute the transversal spin accumulation in normal and superconducting systems.

\begin{figure}
   \begin{center} 
   \includegraphics[width=\columnwidth]{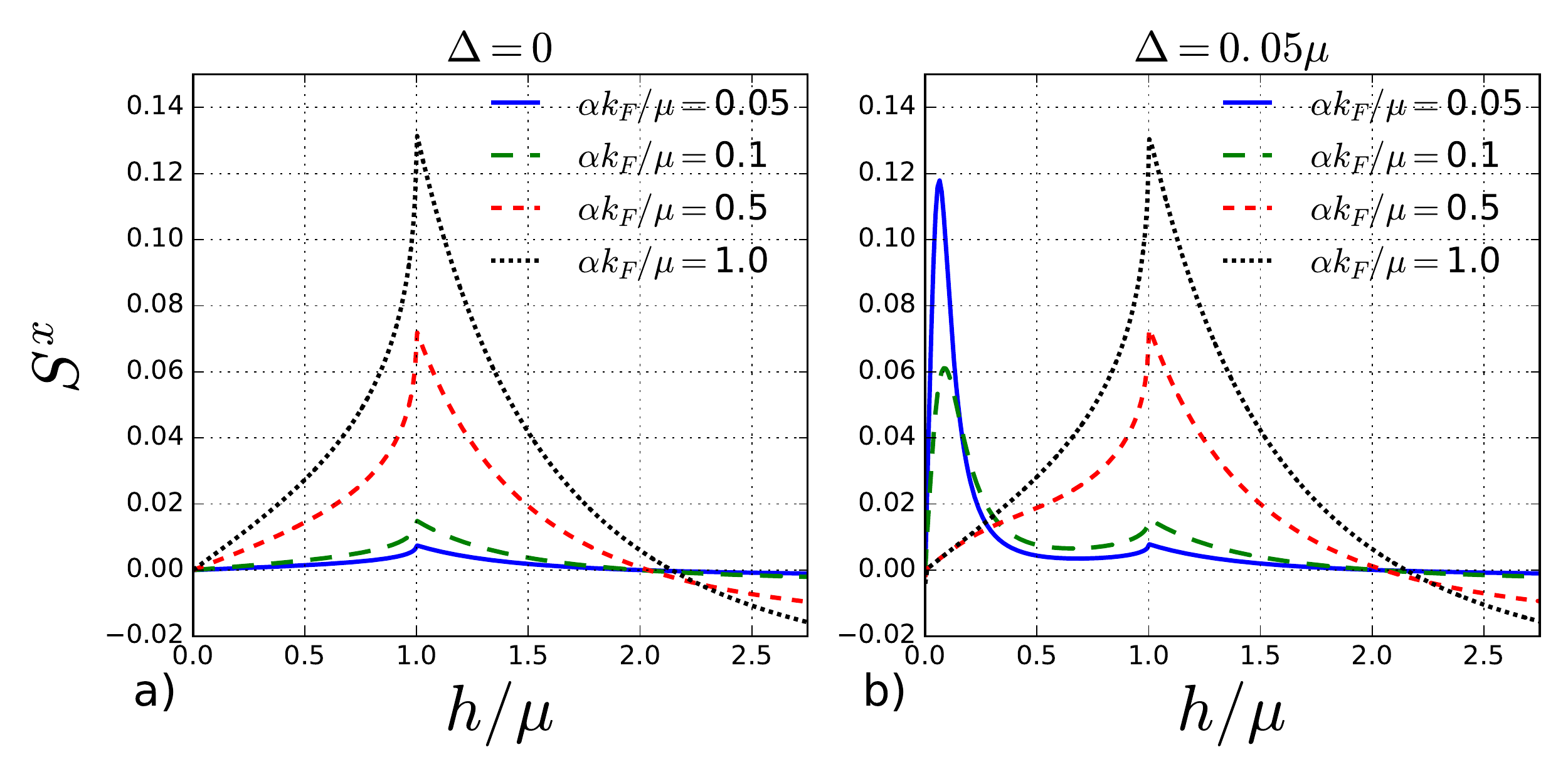}
   \caption{Transverse spin accumulation as function of the Zeeman field for various values of $\alpha k_F$ in the a) normal and b)  superconducting case at $T=0$.
   }\label{Fig:exact}
   \end{center}
\end{figure}

We start with the example of a normal ballistic wire described by the Hamiltonian in Eq.(\ref{Eq01_Hamiltonian}) without interactions ${H}_{int}=V_c=0$ \cite{Dolcini2018}. The spin current can be written as 
 \begin{flalign}
 	J_x^y=\int \frac{dk}{2\pi} \text{Tr}\left[\sigma^y \hat{v}_x\hat{\rho}_k\right]
 \end{flalign}
with the velocity operator $\hat{v}_x=\frac{\partial {H}}{\partial{\hat{p}_x}}=\frac{k}{m}+\alpha\hat{\sigma}^y$
and the equilibrium density matrix, projector on the two eigenstates  $E^\pm_k=\xi_k\pm E_k$
\begin{flalign}
	\hat{\rho}_k=\sum_{i=\pm}\rho^i_k f(E^i_k),\;\rho^\pm_k=\frac{1}{2}\left(\mathds{1}\pm \frac{\alpha k\hat{\sigma}y+h\hat{\sigma}^z}{E_k}\right)
\end{flalign}
respectively, where $\xi_k=k^2/2m-\mu$, $E_k=\sqrt{\alpha^2k^2+h^2}$ and the Fermi distribution is denoted by $f(E)$. We focus on the case $T=0$ and $\mu,\alpha,h>0$. 
Introducing the Fermi momenta for the two spin split bands
\begin{flalign}
	k^{\pm}=\sqrt{2m\left[m\alpha^2+\mu\mp\sqrt{\left(m\alpha^2+\mu\right)^2+h^2-\mu^2}\right]}
\end{flalign}
the current can be expressed as the sum of the contributions from the two spin splitt bands 
\begin{flalign}\label{eq:normal}
	J^y_x=J^{y,-}_x+\Theta(\mu-h)J_x^{y,+}
\end{flalign}
where $+$, $-$ labels the upper, lower band respectively. The contributions are
\begin{flalign}
	J^y_\pm=\frac{\alpha k^\pm}{\pi}\pm\frac{1}{2m\pi}
	\left[
	\frac{k^\pm}{\alpha}\sqrt{(\alpha k^\pm)^2+h^2}\right.\nonumber\\
	\left.-\frac{h^2}{\alpha^2}\ln\left(\sqrt{\frac{(\alpha k^\pm)^2}{h^2} +1}+\frac{\alpha}{h}k^\pm\right)
	\right].\label{Jnormwire}
\end{flalign}
Clearly the ESC is finite only if both the SOC and Zeeman field are finite. The transverse spin accumulation can be computed by substituting Eq.~(\ref{Jnormwire}) into Eq.~(\ref{eq:S_accumulation}). The result is shown in Fig.~\ref{Fig:exact}a. For $h<\mu$ the spin current increases until to point $h = \mu$ where $k^+ = 0$. For larger $h$ the upper band does not contribute, which results in a cusp-like maximum in the spin current, in accordance to the results of Ref. \cite{Dolcini2018} for a normal wire 
\footnote{Notice that the expression (\ref{eq:S_accumulation}) makes a connection between the bulk spin-current and the total transverse spin.  The spatial distribution of the latter has to be computed by solving the full boundary value problem. In a normal wire this was done in Ref.~\cite{Dolcini2018}.}.  

\begin{figure}
   \begin{center} 
   \includegraphics[width=\columnwidth]{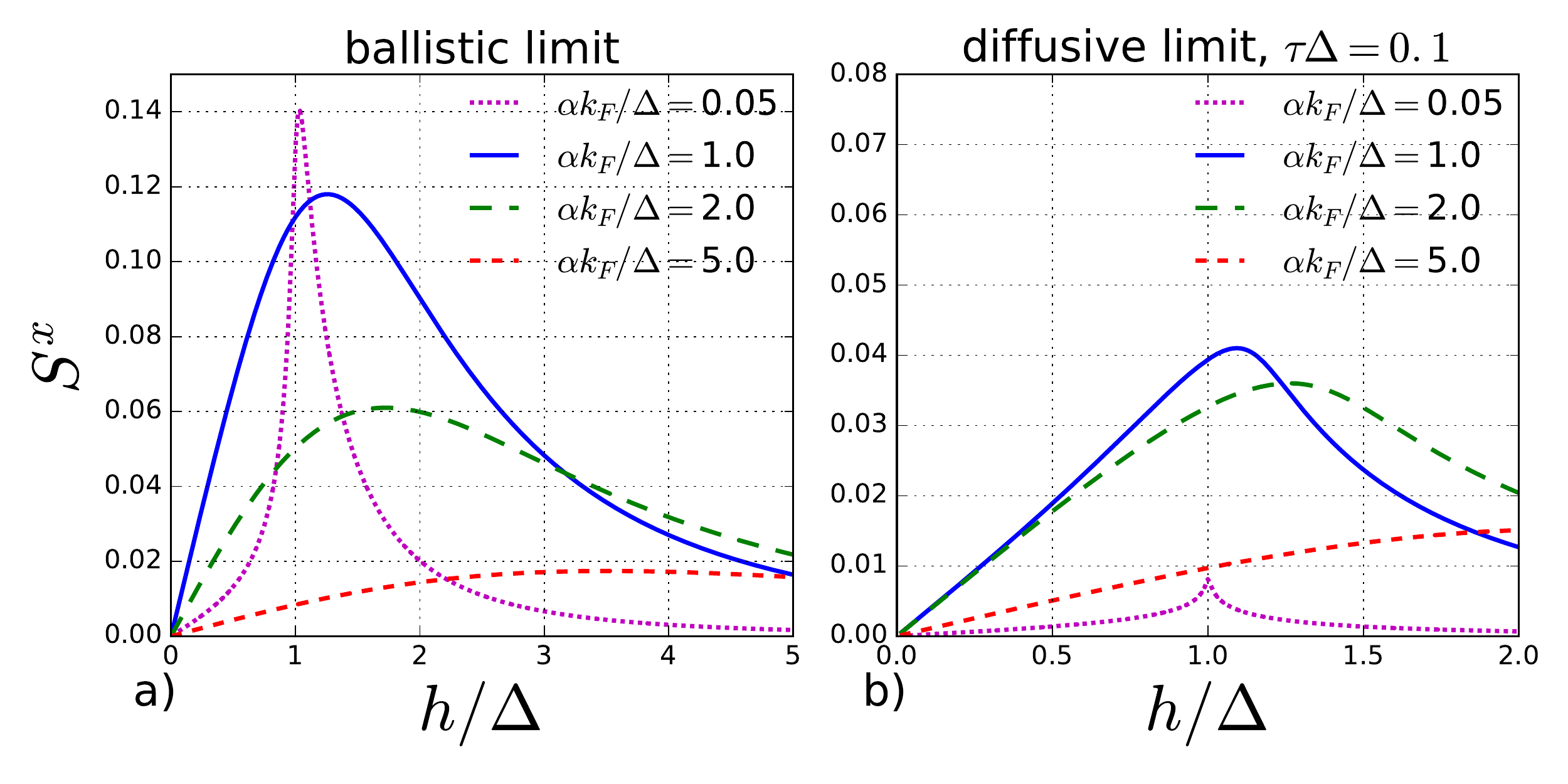}
   \caption{Spin accumulation  calculated in the quasiclassical limit as function of the Zeeman field for various values  of $\alpha k_F$ in the a) ballistic and b) diffusive limit at $T=0$. In the diffusive limit we have chosen $\Delta\tau=0.1$.
   }\label{Fig:classic}
   \end{center}
\end{figure}

Now we assume that the wire is placed on top of a superconductor so that it is fully proximized.  The system is described by the Bogoliubov de Gennes (BdG) Hamiltonian in the Nambu-Spin space
\begin{flalign}\label{eq:SC_Hamiltonian}
	\mathcal{H}=\left(\frac{\hat{p}_x^2}{2m}-\mu+\alpha \hat{p}_x\hat{\sigma}^y\right)\tau_z+h\hat{\sigma}z+\Delta \tau_x
\end{flalign}
where $\Delta$ is the SC order parameter.
The Hamiltonian has the spectrum ($i=1,2$)
\begin{flalign}
	E_{\pm i}=\pm\sqrt{\xi_k^2+(\alpha k)^2+h^2+\Delta^2+(-1)^i 2Q}
\end{flalign}
with $Q=\sqrt{(\xi_k^2+\Delta^2)h^2+\xi_k^2\alpha^2k^2}$.
The spin current can be written as
\begin{flalign}
	J^y_x=\int \frac{dk}{2\pi}\sum_{i=\pm1,\pm2}\text{Tr}\left[P_i\hat{J}^y_x\right]f(E_i)
\end{flalign}
where $P_i=\prod_{j\neq i}\frac{\mathcal{H}-E_j}{E_i-E_j}$ is the projection operator to the eigenstate $|n_i\rangle$ of the BdG Hamiltonian and $\hat{J}^y_x=\left(\frac{k}{m}\hat{\sigma}y+\alpha\right)\tau_z$ the spin current operator in Nambu-spin space. Evaluating the expression at $T=0$ leads to
\begin{flalign}\label{FULL:ballistic}
	&J_x^y=
	\frac{\alpha}{2\pi}
		\int dk
		\left\{	1- \frac{\xi_k}{2}\sum_{i=1,2}
		\left[\frac{1}{E_i}\left(1+(-1)^i\frac{E_k^2}{Q}\right)\right]
		 \right\}\nonumber\\
	&-\frac{\alpha}{2\pi}
	\int dk
	\left\{	 \frac{k^2}{2m}\sum_{i=1,2}\left[\frac{1}{E_i}\left(1+(-1)^i\frac{\xi_k^2}{Q}\right)\right]
	 \right\}.
\end{flalign} The numerical evaluation of the integral shows two prominent features as can be seen in Fig.\ref{Fig:exact}b). The cusp like local maximum also obtained in the normal conducting case is shifted to $h=\sqrt{\mu^2+\Delta^2}$ marking the critical Zeeman field above which the wire is in the topological phase. { The second feature is a peak that appears   within the quasiclassical regime, $h\ll\mu$. For  small enough  values of $\alpha$  this peak is at  $h\approx\Delta$. 
By  lowering $\alpha$ the peak first gets narrow and simultaneously increases in magnitude exceeding  by far the local maximum at the topological transition.  Further decrease of $\alpha$  reduces the height of the peak until the spin current,  and thus the spin accumulation,  vanish non-analytically when  $\alpha\to0$.}  To investigate this non-analyticity in $\alpha$ we turn to the quasiclassical approximation where the spin current is given by \cite{Konschelle2014},
\begin{flalign}
	{J}^a_x=\frac{{i}\pi N_0}{2}T\sum_{\omega_n}\text{Tr}\langle{v}_F\hat{\sigma}^a\check{g}\rangle.
\end{flalign}
Here $\check{g}$ is the quasiclassical Green's function that is obtained from the Gorkov Green's function by the $\xi$-integration \cite{Eilenberger1968} $\check{g}=i/\pi\int d\xi \check{G}(k,\omega_n)$, $\langle\cdots\rangle$ is the average over the Fermi surface, $T$ the temperature, $N_0=1/(2\pi v_F)$ is the 1D density of states at the Fermi level, $\omega_n=(2n+1)\pi T$ are the Matsubara frequencies and $v_F$, $k_F$ denote the Fermi velocity, Fermi momentum respectively. At $T\rightarrow0$ we obtain for the spin current
\begin{flalign}
	J_x^y=\int \frac{d\omega}{\pi} \text{Im}\left[
	\frac{\alpha k_F\left(\omega^2+\Delta^2+iR\right)}{R\sqrt{\omega^2+\Delta^2-h^2-\alpha^2k_F^2+2{i}R}}
	\right]
\end{flalign}
with $R=\sqrt{h^2 \omega^2+(\alpha k_F)^2(\omega^2+\Delta^2)}$. As shown in Fig.\ref{Fig:classic}a the peak  of $S^x$ at $h=\Delta$  within the quasiclassical approximation  shows the same behaviour as in the exact case ({\it cf.} Fig. 1b).  The peak is more pronounced for small values of SOC and  washes out with increasing SOC strength. To describe the behavior at $h=\Delta$ when $\alpha k_F\ll\Delta$ the integral can be approximated by keeping only leading order terms in $\alpha/\Delta$ which gives
\begin{flalign}
	J_x^y\approx-\frac{\alpha k_F}{\pi}\int_0^\infty d\omega \frac{1}{R^{3/2}}\approx-2\sqrt{\frac{\alpha k_F\Delta}{\pi}}\frac{\Gamma(5/4)}{\Gamma(3/4)}
	\label{eq:peak_ball}
\end{flalign}
where in the second step $R\approx\sqrt{(\omega h)^2+(\alpha k_F)^2}$. In a similar fashion we can compute the spin current at the point $h=\Delta$ for finite temperature and $\alpha k_F \ll T\ll\Delta$ which leads to
\begin{flalign}
	J_x^y\approx-\frac{\alpha k_F}{\pi^\frac{3}{2}}\sqrt{\frac{\Delta}{T}}\sum_{n=-\infty}^{\infty}\frac{1}{|2n+1|^\frac{3}{2}}.
\end{flalign}

The next important question is whether this strong enhancement of the spin accumulation for Zeeman energies close to the gap is robust against disorder. To answer this question we calculate the spin current in the diffusive limit as follows  \cite{Tokatly2017}
\begin{flalign}\label{Eq:currentUsadel}
	J_x^y=-i\frac{\pi D}{2}N_0T\sum_{\omega_n}\text{Tr}
	\left\{
	\hat{\sigma}^y\check{\mathfrak{g}}\tilde{\nabla}_x\check{\mathfrak{g}}
	\right\},
\end{flalign}
where $\check{\mathfrak{g}}$ is the quasiclassical Green's function in the diffusive limit, $D=v_F^2\tau$ is the diffusion constant in 1D, $\tau$ is the momentum relaxation time,  $\tilde{\nabla}_x\cdot=\partial_x\cdot-i(\kappa_\alpha/2)\left[\hat{\sigma}^y,\cdot\right]$, and $\kappa_\alpha=2m\alpha$ is the inverse of the spin precession length.
To find $\check{\mathfrak{g}}$ we solve the Usadel equation for a bulk system\cite{Tokatly2017}
\begin{flalign}\label{Eq:Usadel}
	\frac{1}{2}\left[\hat{\Gamma}\check{\mathfrak{g}},\check{\mathfrak{g}}\right]
	-\left[
	\left(\hat{\omega}_n-i h \hat{\sigma}_z\right)\tau_3+\Delta \tau_1,\check{\mathfrak{g}}
	\right]=0
\end{flalign}
where $\hat{\Gamma}\check{\mathfrak{g}}=\left[\hat{\sigma}^y,\left[\hat{\sigma}^y,\check{\mathfrak{g}}\right]\right]/4\tau_\alpha$, and $1/\tau_\alpha=D\kappa_\alpha^2$ is the inverse  Dyakonov-Perel spin relaxation time. The Green's function has to fulfill the  normalization condition $\check{\mathfrak{g}}^2=\mathds{1}$.

The only non-vanishing components of the normal and anomalous quasiclassical Green's functions $\hat{\mathfrak{g}}$ and $\hat{\mathfrak{f}}$ are the singlet and triplet-$z$ components. Thus we make the ansatz
$\check{\mathfrak{g}}=\hat{\mathfrak{g}}\tau_3+\hat{\mathfrak{f}}\tau_1$ with $\hat{\mathfrak{g}}=\mathfrak{g}_0\hat{\sigma}^0+\mathfrak{g}_z \hat{\sigma}^z$ and\;\;$\hat{\mathfrak{f}}=\mathfrak{f}_0\hat{\sigma}^0+\mathfrak{f}_z \hat{\sigma}^z$. For this setup
in equilibrium and for $T=0$ the expression for the spin current, Eq.(\ref{Eq:currentUsadel}), reduces to 
\begin{flalign}\label{Eq:Usadel_integral}
	J_x^y=-\frac{\alpha k_F}{\Delta}\Delta\tau\int\frac{d\omega}{\pi}\left(1-\mathfrak{g}^+\mathfrak{g}^--\mathfrak{f}^+\mathfrak{f}^-\right)
\end{flalign}
with $\mathfrak{g}^\pm=\mathfrak{g}_0\pm \mathfrak{g}_z$, $\mathfrak{f}^\pm=\mathfrak{f}_0\pm \mathfrak{f}_z$.  { Substitution of the  solution of  Eq.~(\ref{Eq:Usadel}) into Eqs. (\ref{Eq:currentUsadel}-\ref{eq:S_accumulation}) leads to the transverse spin accumulation shown in Fig. \ref{Fig:classic}b.  We find the same maximum in $S^x$  at $h\approx\Delta$  for small values of SOC. }  To investigate the behavior in this parameter regime we notice that Eq.~(\ref{Eq:Usadel}) and the condition $\check{\mathfrak{g}}^2=\mathds{1}$ are equivalent to the following system of equations
\begin{flalign}
&\mathfrak{g}^\pm=\frac{\omega \mp({i}h+\mathfrak{g}_z/2\tau_\alpha)}{\sqrt{\left(\omega\mp\left({i}h+\mathfrak{g}_z/2\tau_\alpha\right)\right)^2+\left(\Delta\mp\mathfrak{f}_z/2\tau_\alpha\right)^2}}\\
&\mathfrak{f}^\pm=\frac{\Delta\mp\mathfrak{f}_z/2\tau_\alpha}{\sqrt{\left(\omega \mp\left({i}h+ \mathfrak{g}_z/2\tau_\alpha\right)\right)^2+\left(\Delta\mp \mathfrak{f}_z/2\tau_\alpha\right)^2}}\; .
\end{flalign}
In the limit $\omega\ll1/\tau_\alpha\ll h\sim\Delta$ we can obtain from these equations the Green's functions. Within the logarithmic accuracy the spin current, Eq.(\ref{Eq:Usadel_integral}), is given by
\begin{flalign}\label{Eq:Usadel_integral_approximate}
 	J_x^y\approx{{\frac{2}{3}\frac{\alpha k_F}{\pi}}}\Delta \tau\log\left(\Delta\tau_\alpha\right)\; . 
\end{flalign}
Thus, in the diffusive case the peak in the spin-current at $h=\Delta$ for small values of the SOC is reduced by a factor $\sim \delta\log\delta$, with  $\delta=\Delta\tau\sqrt{\frac{\alpha k_F}{\Delta}}\ll1$, with respect to the pure ballistic case, {\it cf.} Eq. (\ref{eq:peak_ball}). This explains the large  difference between the peaks heights in Figs.  \ref{Fig:classic}a-b, for $k_F\alpha=0.05\Delta$.  Notice however that this  difference  is not that pronounced   for larger values of the SOC.


The general  expression Eq.~(\ref{eq:S_accumulation}) relates the bulk ESC to the transverse spin-polarization, $S^x$ in a finite wire. 
$S^x$ accumulates at the edge  and decays from the edge towards the bulk, but the exact spatial distribution of $S^x$ has to be determined in each particular case. From symmetry arguments it is also clear  that in a wire of finite length $S^x$  at the two edges has opposite  sign (see Fig.\ref{Fig:scheme}a). In the normal state the dependence  $S^x(x)$ was computed numerically in Ref. \cite{Dolcini2018} for a wire connected to two leads and for values of $h$ of the order of $|\mu|$.  Here we focus on the  superconducting state in the relevant  
quasiclassical regime, $h\ll\mu$, for which the Eilenberger equation holds
\begin{flalign}\label{eilenberger}
\pm {v_F} \partial_x \check g_\pm=i\left[\pm {{\alpha k_F}}\hat{\sigma}^y  	+\tau_3 \left(i\omega_n+h\hat{\sigma}^z\right)+i\tau_2\Delta,\check g_\pm\right]\; .
\end{flalign} 
Here $\check g_\pm$ denotes the the GF for both propagation directions $\pm v_F$. We assume 
that the wire extends over the region $x>0$. At the edge, $x=0$ no current flows and the boundary condition imposes that $\check g_+(0)=\check g_- (0)$.  In order to obtain simple analytical results we perform perturbation in $h$ and write the solution of Eq. \ref{eilenberger}
as   $\check g_\pm(x)=\check g_{BCS}+ \delta \check g_\pm(x)$ where $\check g_{BCS}=(\omega_n\tau_3+\Delta\tau_2)/E$ and $E=\sqrt{\omega_n ^2 +\Delta ^2}$.  
The spin density, or magnetic response,   is obtained from 
$\delta M_{j}=\pi iT\mu_BN_{0}\frac{1}{8}\sum_{\omega_{n}}\text{Tr}\left[\tau_{3}\hat{\sigma}^{j}\delta\check{g}^{s}\right]$, where $\delta \check g^s=\delta\check g_++\delta\check g_-$. Specifically, we obtain 
\begin{equation}
\delta \check{g}^{s}=\frac{2h\Delta \check{g}_{BCS}\tau_{1}}{E^{2}+({{\alpha k_F}})^{2}}\left[\hat{\sigma}^{z}+\frac{{{\alpha k_F}}}{E} e^{-\left(i\hat{\sigma}^{y}\kappa_\alpha+\frac{2E}{v_F}\right) x}\right]\label{eq:isot_g0}\; ,
\end{equation}
and hence
\begin{equation}
\frac{M_{x}}{M_0}=\Delta^{2}\pi T\sum_{\omega_n} \frac{{{\alpha k_F}}\cos(\kappa_\alpha x)}{E^{2}\left(E^{2}+({{\alpha k_F}})^{2}\right)}e^{-\frac{2Ex}{v_F}},\label{eq:deltaMx}
\end{equation}
 \begin{equation}
\frac{\delta M_{z}}{M_0}=\Delta^{2}\pi T\sum_{\omega_n} 
\frac{E+{{\alpha k_F}} \sin(\kappa_\alpha x)e^{-\frac{2Ex}{v_F}}}{E^2(E^{2}+({{\alpha k_F}})^{2})}\; , 
\label{eq:deltaMz}
\end{equation}
where  $M_0=\mu_B h N_0$ is the Pauli paramagnetic term. This is a remarkable result  that generalizes the Knight shift 
in superconducting systems with intrinsic SOC \cite{Gorkov2001}. In a finite system  besides the longitudinal response, $M_z=\delta M_z-M_0$, there is a finite transverse magnetization $M_x$ accumulated at the edge of the sample which  decays toward the bulk over the superconducting coherent length.  At low temperatures,   the ratio  $M_x/M_z$ at the edge of the sample, $x=0$,  is proportional to $\Delta/\alpha k_F$ and therefore the transverse magnetization can be much larger than $M_z$ provided $\alpha k_F\ll\Delta$. 
The full spatial dependence of $M_{x,z}(x)$ is shown in  Fig.~\ref{Fig:scheme}b.

\begin{figure}
   \begin{center} 
   a) \;\;\;\;\;\;\;\includegraphics[width=0.6\columnwidth]{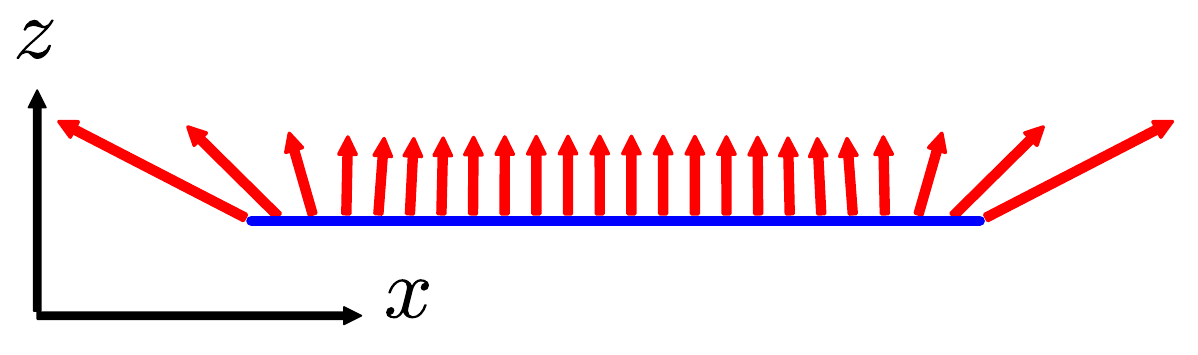}\;\;\;\;\;\;\;\;\;\;\;\;
   b)\includegraphics[width=0.8\columnwidth]{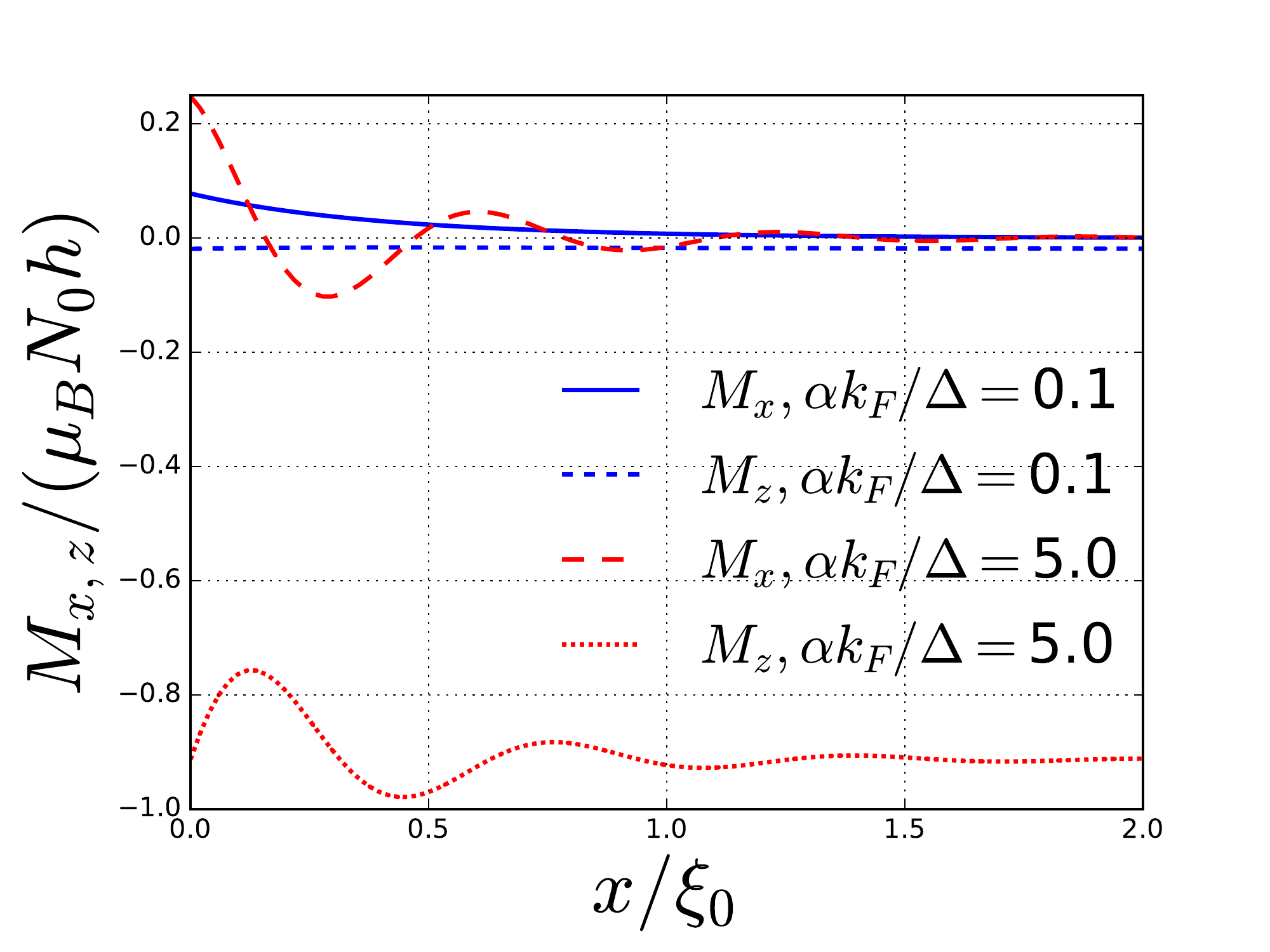}
   \caption{a) Scheme of the total magnetization (red arrows) along a wire of finite length (blue solid line) when a Zeeman field is applied in $z$-direction b) Spatial dependence of the mangetization parallel and perpendicular to the exchange field for a semi-infinite wire and different values of SOC. In all plots $T=0$. }\label{Fig:scheme}
   \end{center}
\end{figure}

In summary, we have demonstrated a universal correspondence between equilibrium spin currents and a transverse 
spin accumulation in wires with SOC and a perpendicular Zeeman field. Such spin accumulation appears in both normal and superconducting systems. In the normal state the total edge spin is maximized for values of $h$ of the order of the chemical potential $\mu$. More interesting, in the superconducting state the effect can be maximized at values of the SOC and Zeeman energy much smaller than $\mu$. We demonstrated that this  effect is robust against disorder and manifest in a hitherto  unknown transverse magnetic susceptibility in finite systems.

\bigskip

{\it Acknowledgments.-} We acknowledge funding by the Spanish Ministerio de Ciencia, Innovacion y Universidades (MICINN) (Projects No.  FIS2016- 79464-P and FIS2017-82804-P). I.V.T. acknowledges support by Grupos Consolidados UPV/EHU del Gobierno Vasco (Grant No. IT578-13). F.S.B. and B. B acknowledge funding from the EU's Horizon 2020  and innovation programme under grant agreement No. 800923 (SUPERTED).

\bibliographystyle{new}
\bibliography{library}

\begin{thebibliography}{10}

\bibitem{Fabian2004}
I.~{\v{Z}}uti{\'{c}}, J.~Fabian, and S.~{Das Sarma},
\newblock Rev. Mod. Phys. {\bf 76}, 323 (2004).

\bibitem{Rashba2007}
E.~I. Rashba,
\newblock {in {\it Future Trends in Microelectronics: Up the Nano Creek},
  edited by S. Luryi, J. M. Xu, and A. Zaslavsky (Wiley, Hoboken, 2007).}

\bibitem{Maekawa2012}
S.~Maekawa, S.~O. Valenzuela, E.~Saitoh, and T.~Kimura, editors,
\newblock {\em {Spin Current}} (Oxford University Press, 2012).

\bibitem{Eschrig2011}
M.~Eschrig,
\newblock Phys. Today {\bf 64}, 43 (2011).

\bibitem{Shi2006}
J.~Shi, P.~Zhang, D.~Xiao, and Q.~Niu,
\newblock Phys. Rev. Lett. {\bf 96}, 076604 (2005).

\bibitem{EngelRashbaHalperin2007}
H.-A. Engel, E.~I. Rashba, and B.~I. Halperin,
\newblock {in {\it Handbook of Magnetism and Advanced Magnetic Materials},
  edited by H. Kronm\"uller and S. Parkin (Wiley, New York, 2007), Vol. 5.}

\bibitem{Rashba2003}
E.~I. Rashba,
\newblock Phys. Rev. B {\bf 68}, 241315 (2003).

\bibitem{Tokatly2008}
I.~V. Tokatly,
\newblock Phys. Rev. Lett. {\bf 101}, 106601 (2008).

\bibitem{Tokatly2010}
I.~V. Tokatly and E.~Sherman,
\newblock Annals of Physics {\bf 325}, 1104 (2010).

\bibitem{Sonin2007a}
E.~B. Sonin,
\newblock Phys. Rev. B {\bf 76}, 033306 (2007).

\bibitem{Sonin2007}
E.~B. Sonin,
\newblock Phys. Rev. Lett. {\bf 99}, 266602 (2007).

\bibitem{Dolcini2018}
F.~Dolcini and F.~Rossi,
\newblock Phys. Rev. B {\bf 98}, 045436 (2018).

\bibitem{abrikosov1962spin}
A.~A. Abrikosov and L.~P. Gor'kov,
\newblock Sov. Phys. JETP. {\bf 15}, 752 (1962).

\bibitem{Abrikosov1988}
A.~A. Abrikosov,
\newblock {\em {Fundamentals of the Theory of Metals}} (North Holland, 1988).

\bibitem{Gorkov2001}
L.~P. Gor'kov and E.~I. Rashba,
\newblock Phys. Rev. Lett. {\bf 87}, 037004 (2001).

\bibitem{Sau2010}
J.~D. Sau, R.~M. Lutchyn, S.~Tewari, and S.~{Das Sarma},
\newblock Phys. Rev. Lett. {\bf 104}, 040502 (2010).

\bibitem{Oreg2010}
Y.~Oreg, G.~Refael, and F.~{Von Oppen},
\newblock Phys. Rev. Lett. {\bf 105}, 177002 (2010).

\bibitem{Mourik2012}
V.~Mourik {\em et~al.},
\newblock Science {\bf 336}, 1003 (2012).

\bibitem{Das2012}
A.~Das {\em et~al.},
\newblock Nature Physics {\bf 8}, 887 (2012).

\bibitem{Deng2012}
M.~T. Deng {\em et~al.},
\newblock Nano Lett. {\bf 12}, 6414 (2012).

\bibitem{Hansen2016}
M.~T. Deng {\em et~al.},
\newblock Science {\bf 354}, 1557 (2016).

\bibitem{Sticlet2012a}
D.~Sticlet, C.~Bena, and P.~Simon,
\newblock Phys. Rev. Lett. {\bf 108}, 096802 (2012).

\bibitem{Ojanen2012}
T.~Ojanen,
\newblock Phys. Rev. Lett. {\bf 109}, 226804 (2012).

\bibitem{Szumniak2017}
P.~Szumniak, D.~Chevallier, D.~Loss, and J.~Klinovaja,
\newblock Phys. Rev. B {\bf 96}, 041401 (2017).

\bibitem{Bjornson2015}
K.~Bj{\"{o}}rnson, S.~S. Pershoguba, A.~V. Balatsky, and A.~M. Black-Schaffer,
\newblock Phys. Rev. B {\bf 92}, 214501 (2015).

\bibitem{Note1}
Notice that the expression (\ref {eq:S_accumulation}) makes a connection
  between the bulk spin-current and the total transverse spin. The spatial
  distribution of the latter has to be computed by solving the full boundary
  value problem. In a normal wire this was done in Ref.~\cite {Dolcini2018}.

\bibitem{Konschelle2014}
F.~Konschelle,
\newblock European Physical Journal B {\bf 87} (2014).

\bibitem{Eilenberger1968}
G.~Eilenberger,
\newblock Zeitschrift f{\"u}r Physik {\bf 214}, 195 (1968).

\bibitem{Tokatly2017}
I.~V. Tokatly,
\newblock Phys. Rev. B {\bf 96}, 060502 (2017).

\end{thebibliography}

\end{document}